\pdfoutput=1

\documentclass[11pt]{article}

\usepackage[final]{acl}
\usepackage{times}
\usepackage{latexsym}
\usepackage[T1]{fontenc}            % allow accented characters
\usepackage[utf8]{inputenc}
\usepackage{microtype}              % improve spacing
\usepackage{inconsolata}            % nicer tt font
\usepackage{graphicx}               % include graphics
\usepackage{booktabs}               % nice tables
\usepackage{comment}
\usepackage{float}
\usepackage{pgfplots}
\usepackage{array}
\usepackage{tabularx}
\usepackage{enumitem}
\usepackage{multirow}
\usepackage{hyperref}
\newcolumntype{Y}{>{\raggedright\arraybackslash}X} % flexible, left-aligned col
\newcolumntype{R}{>{\raggedleft\arraybackslash}X} % flexible right-aligned column

\usepackage{listings}
\usepackage{xcolor}
\usepackage{upquote} % straight quotes in typewriter fonts
\lstdefinestyle{promptjson}{
  basicstyle=\ttfamily\footnotesize,
  columns=fixed,
  breaklines=true,
  keepspaces=true,
  showstringspaces=false,
  frame=single,
  framerule=0.4pt,
  backgroundcolor=\color{black!3},
  aboveskip=6pt, belowskip=6pt
}
\lstdefinestyle{prompttext}{
  basicstyle=\ttfamily\footnotesize,
  columns=fullflexible,
  breaklines=true,
  showstringspaces=false,
  frame=single,
  framerule=0.4pt,
  backgroundcolor=\color{black!3},
  aboveskip=6pt, belowskip=6pt
}

\title{Are LLMs Court-Ready? Evaluating Frontier Models on Indian Legal Reasoning}

\author{
  Kush Juvekar\\
  \texttt{kush@adalat.ai} \\
  {\normalfont Adalat AI, India}
  \And
  Arghya Bhattacharya\\
  \texttt{arghya@adalat.ai} \\
  {\normalfont Adalat AI, India}
  \And
  Sai Khadloya \\
  \texttt{sai@adalat.ai} \\
  {\normalfont Adalat AI, India}
  \And
  Utkarsh Saxena \\
  \texttt{utkarsh@adalat.ai} \\
  {\normalfont Adalat AI, India}
}

\begin{document}
\maketitle

% ===================== ABSTRACT (≤150 words) =====================
\begin{abstract}
Large Language Models are entering legal workflows, yet we lack a jurisdiction-specific framework to assess their baseline competence therein. We use India’s public legal examinations as a transparent proxy. Our multi-year benchmark assembles objective screens from top national and state exams and evaluates open and frontier LLMs under \emph{real world exam conditions}. To probe beyond MCQs, we also include a lawyer-graded, paired-blinded study of long-form answers from the Supreme Court’s Advocate-on-Record exam. This is, to our knowledge, the first exam-grounded, India-specific yardstick for LLM court-readiness released with datasets and protocols. Our work shows that while frontier systems consistently clear historical cutoffs and often match or exceed recent top-scorer bands on objective exams, none surpasses the human topper on long-form reasoning. Grader notes converge on three reliability failure modes—procedural/format compliance, authority/citation discipline, and forum-appropriate voice/structure. These findings delineate where LLMs can assist (checks, cross-statute consistency, statute and precedent lookups) and where human leadership remains essential: forum-specific drafting and filing, procedural and relief strategy, reconciling authorities and exceptions, and ethical, accountable judgment.

\end{abstract}

% ===================== 1. INTRODUCTION (short) =====================
\section{Introduction}
LLMs have cleared multiple-choice bar-style screens in several jurisdictions, renewing interest in legal automation, but a jurisdiction-first question remains: \emph{are these systems court ready}? Other fields probe such capability with exam-style settings: broad knowledge suites such as MMLU, Olympiad-level problems in mathematics and science, and clinically oriented reasoning in health \citep{hendrycks2020mmlu,olympiadbench_acl2024,multimedqa_nature}. By contrast, many AI-and-law studies focus on short-context recall (for example, bail or recidivism prediction and legal-judgment prediction). These are metric-friendly but only indirectly tied to how courts expect lawyers to write and file \citep{kleinberg2018human,dressel2018recidivism,cui2022ljp_survey}. India offers a jurisdiction where court-legible benchmarks already exist. We adopt \emph{public exams} already used to gate human entry \textbf{Common Law Admission Test (CLAT) –UG/PG} (admissions), \textbf{Delhi Judicial Services/ Delhi Higher Judicial Services (DJS/DHJS)} prelims (judiciary), and the Supreme Court’s \textbf{Advocate-on-Record (AoR)} exam (rights of audience) as court-ready yardsticks \citep{clat_ug_syllabus_2026,clat_pg_pattern,djs_instr_2023,dhjse_imp_instr,aor_sci_official}.

\vspace{0.2cm}
\textbf{Our primary contributions in this paper are:}
\vspace{0.2cm}
\begin{itemize}[leftmargin=*, itemsep=0.3ex, topsep=0.4ex, parsep=0pt, partopsep=0pt]
  \item \textbf{Exam-grounded dataset (objective + subjective):} We curate a multi-year corpus of \emph{objective} questions \textbf{6{,}218 MCQs} plus \emph{subjective} AoR materials (2023). Provenance, year coverage, and marking rules are documented in the Appendix. We release the dataset \href{https://huggingface.co/datasets/adalat-ai/indian-legal-exam-benchmark}{here.}\raisebox{-0.2\height}{\includegraphics[height=1.6em]{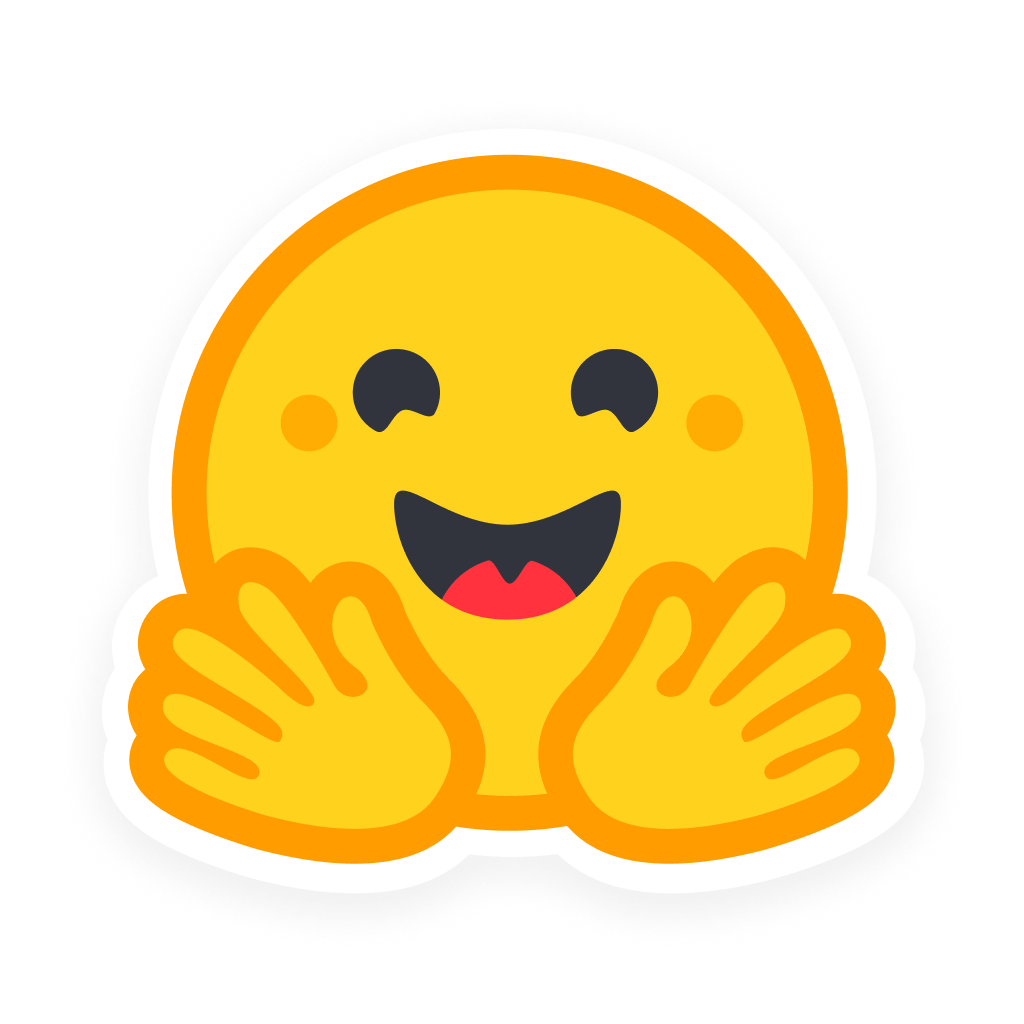}}
  \item \textbf{Benchmark under official rules:} We evaluate open and closed models including frontier and strong open baselines under exam-native interfaces and identical scoring policies, enabling comparisons across model families and scales.
  \item \textbf{Blinded AoR study with certified graders.} For each AoR paper, we create paired sets comparing the human-written version with the model-generated version. We anonymize them and have certified AoRs evaluate them using a rubric. 
\end{itemize}

\noindent By anchoring evaluation in public exams that every law student, judge, and policymaker recognizes, we present results that legal practitioners can interpret and ML researchers can reproduce. We hope this shared yardstick helps both communities see where current models stand today and guides evidence-based adoption and research.

% ===================== 2. RELATED WORK (short) =====================
\section{Related Work}
Legal NLP suites such as LexGLUE and LegalBench cover broad tasks, and IL–TUR targets Indian legal texts, but none align with the public exams that govern entry and practice in India \citep{lexglue_acl,legalbench_arxiv,iltur_acl}. Outside law, exam-style evaluations (e.g., MMLU and Olympiad-level benchmarks) stress reasoning but do not test forum-specific procedure or authority discipline \citep{hendrycks2020mmlu,olympiadbench_acl2024,olymmath_2025}. In AI and law, much of the literature concentrates on economically salient prediction tasks (for example, bail, recidivism, and legal-judgment prediction), where metrics are tractable but only weakly aligned with the linguistic and rhetorical demands of courtroom writing \citep{kleinberg2018human,dressel2018recidivism,cui2022ljp_survey,shui2023llm_ljp}. Bar-exam studies are informative but jurisdictionally distinct from India’s public exams \citep{katz2023gpt4bar}. To our knowledge, this is the first study to evaluate frontier LLMs on India’s public legal examinations, pairing multi-year objective screens with a lawyer-graded subjective study under exam constraints.

\section{Exams and Scope}
\label{sec:exams}

\paragraph{Why these exams?}
India’s legal profession is structured around a publicly administered system of entry and advancement that is mediated through a series of high-stakes examinations. These include: (i) the CLAT–UG/PG for entry into undergraduate and postgraduate law programs, (ii) judicial service examinations such as the DJS and DHJS for recruitment to the judiciary, and (iii) the Advocate-on-Record (AoR) examination, which confers exclusive rights of audience before the Supreme Court of India. Reliance on these standardized assessments provides institutional legibility for stakeholders, who use them to regulate access, allocate professional privileges, and validate competence across the legal system for humans. \citep{clat_ug_syllabus_2026,clat_pg_pattern,djs_instr_2023,dhjse_imp_instr,aor_sci_official}.

\paragraph{Why AoR for the subjective study?}
The Supreme Court’s \emph{Advocate-on-Record (AoR)} certification is uniquely consequential: under the \emph{Supreme Court Rules, 2013}, only an AoR may file an appearance or act for a party in the Supreme Court (the AoR is the filing advocate of record) \citep{scrules2013}. Eligibility itself signals seniority and training (four years’ practice plus one year of training under an AoR, followed by a Court-conducted examination) \citep{aor_sci_official,aor_training_notice}. The exam is administered by the Supreme Court and consists of four descriptive papers \emph{Practice \& Procedure}, \emph{Drafting}, \emph{Advocacy/Professional Ethics}, and \emph{Leading Cases} making it the most premium, publicly administered subjective legal reasoning assessment tied directly to Supreme Court practice \citep{aor_sci_official,aor_syllabus_overview}. These features make AoR the clearest lens for open-ended legal reasoning and writing.

\paragraph{On excluding AoR \emph{Drafting}.}
We exclude Paper II \emph{(Drafting)} from quantitative scoring because drafting in the AoR exam is format critical: cause-title, parties, prayers, affidavits, layout, and citations must follow strict Supreme Court templates. Without document-template tooling, LLMs generate legally plausible text that routinely violates these formal requirements. In a pilot using Gemini 2.5 Pro, the certified examiner deemed the draft "not evaluable" due to pervasive structural non-compliance (see Appendix \S\ref{app:aor_drafting_note}). As Gemini 2.5 Pro is the only model with consistent cross-exam performance across years, this single pilot suffices to indicate drafting limitations. Accordingly, we evaluate AoR on the three papers testing legal reasoning: \emph{Practice \& Procedure}, \emph{Advocacy \& Professional Ethics}, and \emph{Leading Cases}. 

\paragraph{Year selection and provenance.}
For \emph{objective} exams, we include only years where the governing body released both the official paper and the \emph{official answer key}; years without a key are excluded. For \emph{subjective}, we evaluate AoR~2023 with paired, blinded grading by certified practitioners (three papers as above). All artifacts are sourced from official portals.\footnote{Representative sources: \citet{clat_ug_syllabus_2026,clat_pg_pattern,djs_instr_2023,dhjse_imp_instr,aor_sci_official}.}
A compact dataset summary is in Table \ref{tab:exam-totals-main}; full syllabi and exact year coverage appear in Tables \ref{tab:appendix-syllabus} and \ref{tab:appendix-years}.

\begin{table}[!t]
\centering
\footnotesize
\setlength{\tabcolsep}{3pt}\renewcommand{\arraystretch}{1.05}
\begin{tabularx}{\columnwidth}{l c c Y R}
\toprule
\textbf{Exam} & \textbf{Mod.} & \textbf{Qs/exam} & \textbf{Marking} & \textbf{Total Questions} \\
\midrule
CLAT UG       & Obj.  & 200/150/120 & $+1/-0.25/0$                 & \textbf{3{,}154} \\
CLAT PG       & Obj.  & 120 & $+1/-0.25/0$                 & \textbf{814} \\
DJS (Prelim.)  & Obj.  & 200 & $+1/-0.25$                   & \textbf{1{,}400} \\
DHJS (Prelim.) & Obj.  & 150 & $+1/-0.25$                   & \textbf{850} \\
AoR 2023 (SC)       & Subj. & 35 & none (per-question marks)    & \textbf{35} \\
\bottomrule
\end{tabularx}
\caption{Objective exam sizes and marking (official).}
\label{tab:exam-totals-main}
\end{table}

% ===================== 4. MODELS AND INFERENCE =====================
\section{Model Selection and Inference Configuration}

\paragraph{Objective–exam cohort.}
We evaluate a broad panel spanning frontier proprietary, large open, and strong small/open baselines. The cohort (i) covers \textbf{families} across vendors (Google, OpenAI, Anthropic, Mistral, DeepSeek, Alibaba/Qwen, Meta) to reduce recipe bias; (ii) spans \textbf{scales} from $\sim$7B to frontier sizes to observe size trends under identical constraints; (iii) includes both instruction-tuned and \textbf{reasoning-tuned} models (e.g., R1) to test whether explicit reasoning helps under exam conditions; and (iv) uses widely available endpoints, increasing reproducibility.

\paragraph{Subjective–exam cohort (AoR).}
Human grading constraints preclude evaluating all models on long-form papers. We therefore select a principled triad: 
\emph{Gemini~2.5~Pro} (the objective leader, testing transfer from capacity to drafting), 
\emph{Gemma~3~27B} (the strongest small/open baseline in our objective runs, offering cost-efficient human evaluation), 
and \emph{DeepSeek~R1} (a large, reasoning-tuned open model, probing whether reasoning training enhances forum-specific drafting). 
Together, these models represent a \textbf{frontier ceiling}, a \textbf{competitive small/open baseline}, and a \textbf{reasoning-tuned large open} reference point, sufficient to expose transfer gaps without exhausting grader bandwidth.

\paragraph{Inference setup.}
We replicate \emph{official exam conditions} by incorporating constraints (such as negative marking for MCQs and forum-specific AoR instructions) into our prompts, positioning each model as an exam candidate. Full prompt text is in Appendix~\S\ref{app:prompts_obj} and \S\ref{app:prompts_aor}. Decoding is deterministic (temperature = 0; default top-$p$); no tools, no retrieval, single-pass inference. All endpoints are invoked via a single gateway \emph{OpenRouter} \citep{OpenRouter2025} with date-pinned model identifiers; total evaluation cost was $<\$\,500$. For objective papers we enforce a structured output schema and score strictly under official rules; per-model breakdowns appear in the Appendix\S\ref{app:results}.

% ===================== 5. EVALUATION =====================
\section{Evaluation}

\subsection{Objective exams}
We score per question using the official schema. Models return strict JSON with a single \texttt{answer\_label} using structured outputs; non-conforming outputs are marked wrong in the conservative variant.

\subsection{Subjective exams}

\smallskip
\noindent\textbf{Paired, blinded design.}
For AoR~2023 we obtain human answer script (\textbf{\emph{AoR 2023 Exam Topper}}) and generate one LLM script per model in our subjective cohort (Gemini~2.5~Pro; Gemma~3~27B; DeepSeek~R1).
For each paper, we create three \emph{paired sets}, each containing two anonymized scripts of the same question(s): \emph{Script~A} (human or LLM) and \emph{Script~B} (the other), with order randomized.                    
\textbf{Rubric and aggregation.} Each certified \emph{Advocate-on-Record} (AoR) grader receives all three paired sets and is not told which script is human/AI. Five certified AoR assessors grade all pairs using an official‑style rubric covering: (i) Accuracy and application of law to facts, (ii) Authority discipline (presence, correctness, and fit of case/statute citations; penalties for fabrication/miscitation; rewards for pin‑point cites), (iii) Forum‑specific structure and procedure (orders/rules, cause‑titles/party arrays/prayers where relevant), (iv) Depth/nuance and handling of counter‑arguments, (v) Language and expression (clarity, concision, tone). We compute per-paper totals, deltas (LLM vs. human), and summarize qualitative failures.

\begin{table*}[t]
\centering
\scriptsize % Changed from \footnotesize to \scriptsize for smaller text
\setlength{\tabcolsep}{3pt} % Reduced from 5pt to 3pt to save space
\renewcommand{\arraystretch}{1.0} % Reduced from 1.05 to make rows tighter
\begin{tabularx}{\textwidth}{l l *{10}{>{\centering\arraybackslash}X}} % Use X columns for auto-width adjustment
\toprule
 &  & \multicolumn{2}{c}{\textbf{Grader 1 (AoR)}} & \multicolumn{2}{c}{\textbf{Grader 2 (AoR)}} & \multicolumn{2}{c}{\textbf{Grader 3 (AoR)}} & \multicolumn{2}{c}{\textbf{Grader 4 (AoR)}} & \multicolumn{2}{c}{\textbf{Grader  5 (AoR)}} \\
\cmidrule(lr){3-4}\cmidrule(lr){5-6}\cmidrule(lr){7-8}\cmidrule(lr){9-10}\cmidrule(lr){11-12}
\textbf{Model} & \textbf{Paper} & \textbf{AI} & \textbf{Human} & \textbf{AI} & \textbf{Human} & \textbf{AI} & \textbf{Human} & \textbf{AI} & \textbf{Human} & \textbf{AI} & \textbf{Human} \\
\midrule
\addlinespace[0.1ex] % Reduced from 0.2ex
\textbf{Gemini 2.5 Pro} & Practice \& Proc.        & 63.5 & 73.0 &  68.5  &  69  &  68  &  98  &  72  &  85  &  70  &  83  \\
                        & Adv. \& Ethics           & 79.0 & 78.5 &  72  &  70.5  &  68  &  70  &  68  &  74  &  73  &  69  \\
                        & Leading Cases            & 73.0 & 73.0 &  63  &  75  &  58  &  76  &  58  &  73  &  60  &  75  \\
                        & \textbf{Non-draft. total} & \textbf{215.5} & \textbf{224.5} &  \textbf{203.5} &  \textbf{214.5} &  \textbf{194} &  \textbf{244} &  \textbf{198} &  \textbf{232} &  \textbf{203} &  \textbf{227} \\
\addlinespace[0.2ex] % Reduced from 0.4ex
\textbf{DeepSeek R1}    & Practice \& Proc.        & 67.0 & 73.0 &  59.5  &  69  &  58  &  98  &  50  &  85  &  63  &  85  \\
                        & Adv. \& Ethics           & 68.0 & 78.5 &  64.5  &  70.5  &  63  &  70  &  60  &  74  &  66  &  69  \\
                        & Leading Cases            & 59.0 & 73.0 &  45.5  &  75 &  42  &  76  &  38  &  73  & 43   &  75  \\
                        & \textbf{Non-draft. total} & \textbf{194.0} & \textbf{224.5} &  \textbf{169.5} &  \textbf{214.5} &  \textbf{163} &  \textbf{244} &  \textbf{148} & \textbf{232} &  \textbf{172} &  \textbf{227} \\
\addlinespace[0.2ex]
\textbf{Gemma 3 27B}    & Practice \& Proc.        & 32.0 & 73.0 & 57.5 & 69.0 &  37  &  98  &  41  &  85  &  39  &  85  \\
                        & Adv. \& Ethics           & 58.0 & 78.5 & 41.5 & 70.5 &  66  &  70  &  66  &  74  &  64  &  69  \\
                        & Leading Cases            & 57.0 & 73.0 & 55.0 & 75.0 &  47  &  76  & 47  &  73  &  50  &  75  \\
                        & \textbf{Non-draft. total} & \textbf{147.0} & \textbf{224.5} & \textbf{154.0} & \textbf{214.5} &  \textbf{150} &  \textbf{244} &  \textbf{148} &  \textbf{232} &  \textbf{153} &  \textbf{227} \\
\bottomrule
-\end{tabularx}
\caption{AoR 2023 (paired, blinded): combined per-paper scores by evaluator. Drafting excluded from aggregates; pilot note in Appendix~\S\ref{app:aor_drafting_note}.}
\label{tab:aor-merged}
\end{table*}

% ===================== 6. RESULTS =====================

\begin{table*}[t]
\centering
\small
\setlength{\tabcolsep}{5.5pt}\renewcommand{\arraystretch}{1.08}
\begin{tabularx}{\textwidth}{l *{8}{>{\raggedleft\arraybackslash}X}}
\toprule
\textbf{Model} &
\multicolumn{2}{c}{\textbf{CLAT UG} (/200/150/120)} &
\multicolumn{2}{c}{\textbf{CLAT PG} (/120)} &
\multicolumn{2}{c}{\textbf{DJS} (/200)} &
\multicolumn{2}{c}{\textbf{DHJS} (/150/125)} \\
\cmidrule(lr){2-3}\cmidrule(lr){4-5}\cmidrule(lr){6-7}\cmidrule(lr){8-9}
& \textbf{Avg} & \textbf{$\Delta$} & \textbf{Avg} & \textbf{$\Delta$} & \textbf{Avg} & \textbf{$\Delta$} & \textbf{Avg} & \textbf{$\Delta$} \\
\midrule
\multicolumn{1}{l}{\textit{Topper average (anchor)}} & 148.9 &   & 88.6 &   & 162.7 &   & 114.1 &   \\
\midrule
Gemini 2.5 Pro       & 156.6 & $+7.6$  & 102.8 & $+14.3$ & 182.5 & $+19.8$ & 139.4 & $+25.3$ \\
GPT 5 Chat          & 134.7 & $-14.3$ & 100.8 & $+12.2$ & 162.7 & $+0.0$  & 119.4 & $+5.3$  \\
DeepSeek R1          & 141.6 & $-7.3$  & 91.8  & $+3.2$  & 146.7 & $-16.0$ & 113.3 & $-0.9$  \\
DeepSeek Chat v3     & 141.5 & $-7.4$  & 91.0  & $+2.5$  & 137.3 & $-25.4$ & 114.1 & $+0.0$  \\
Claude Sonnet 4      & 143.5 & $-5.4$  & 88.5  & $-0.1$  & 150.7 & $-12.0$ & 112.8 & $-1.3$  \\
Mistral Medium 3.1   & 136.7 & $-12.2$ & 89.9  & $+1.3$  & 148.1 & $-14.5$ & 106.0 & $-8.1$  \\
Qwen 3 235B          & 138.1 & $-10.9$ & 82.4  & $-6.2$  & 127.8 & $-34.9$ & 83.1  & $-31.0$ \\
Llama 3.3 70B        & 123.0 & $-25.9$ & 82.3  & $-6.3$  & 119.0 & $-43.7$ & 92.5  & $-21.6$ \\
GPT 4.1 Mini         & 122.6 & $-26.3$ & 78.3  & $-10.3$ & 121.9 & $-40.8$ & 88.4  & $-25.7$ \\
Gemma 3 27B          & 117.5 & $-31.4$ & 69.8  & $-18.8$ & 111.3 & $-51.4$ & 80.6  & $-33.5$ \\
Qwen 2.5 7B          & 101.2 & $-47.7$ & 59.7  & $-28.9$ & 76.9  & $-85.8$ & 56.9  & $-57.3$ \\
Gemma 3 12B          & 106.0 & $-42.9$ & 62.3  & $-26.3$ & 94.0  & $-68.7$ & 75.3  & $-38.8$ \\
GPT 3.5 Turbo        & 100.6 & $-48.3$ & 63.7  & $-24.9$ & 85.9  & $-76.8$ & 62.5  & $-51.6$ \\
Llama 3.1 8B         &  90.5 & $-58.4$ & 58.2  & $-30.4$ & 82.1  & $-80.5$ & 54.5  & $-59.6$ \\
\bottomrule
\end{tabularx}
\caption{Cross-exam summary (averages across years). $\Delta$ is against each exam's topper average. Positive $\Delta$ indicates model means at or above topper average. Per-year $\times$ model matrices appear in the Appendix~\S\ref{app:results}}
\label{tab:objective-cross}
\end{table*}

\section{Results}
\subsection{Objective exams}
We evaluate the models against human topper scores released by the official exam committees\cite{clat_2026_site,dhjs_rules_2025}. Across all four objective exams, frontier systems lead consistently. Gemini~2.5~Pro exceeds the historical topper anchors on every exam (e.g., CLAT PG $+14.3$, DJS $+19.8$, DHJS $+25.3$ on average across years), while GPT 5~Chat is near parity on DJS and strongly positive on CLAT PG. Open reasoning-tuned DeepSeek~R1 is competitive on CLAT PG/DHJS but trails on DJS; instruction-tuned DeepSeek~v3 reaches parity on DHJS and small positive on CLAT PG, yet lags on DJS/CLAT UG. Smaller ($\le$30B) models fall below topper anchors across the board (e.g., Gemma~3~27B: $-18.8$ on CLAT PG; $-51.4$ on DJS). 

\subsection{Subjective exams}
Gemini 2.5 Pro demonstrates performance closest to human-level proficiency, achieving near parity on the \emph{Ethics} paper and a statistical tie on \emph{Leading Cases}. However, a notable performance gap remains in the \emph{Practice \& Procedure} paper, suggesting that procedural knowledge and its application present a distinct challenge. In contrast, other models such as DeepSeek R1 and Gemma 3 27B exhibit more significant performance differentials across all examination papers. A qualitative analysis of grader feedback, which proved remarkably consistent across different evaluators, converged on three principal failure modes:

\begin{itemize}[leftmargin=*, itemsep=0.25ex, topsep=0.2ex]
  \item \textbf{Deficiencies in Authority Discipline and Doctrinal Rigor:} A critical shortfall identified across multiple models was the inability to consistently adhere to the conventions of legal citation and authority. This manifested in several forms: the complete omission of controlling precedents; the miscitation of peripheral authorities; and a more subtle failure termed "manufacturing authorities," where a model correctly recalls a relevant case but fails to articulate its specific relevance to the question's legal dilemma. For example, the failure to cite the controlling precedent in a given area, such as omission of \emph{Rupa Ashok Hurra v. Ashok Hurra AIR 2022} in a discussion on curative petitions is not a simple mistake. Furthermore, responses often exhibited a tendency towards "generic" legal assertions - mentioning concepts like "the principles of natural justice" or citing a well-known case like \emph{Maneka Gandhi v.  Union of India AIR 1978} without providing the requisite relevance to the question it is answering. This lack of precision, whether through omission, misapplication, or inadequate synthesis, strips the legal argument of its persuasive force and demonstrates a failure to engage with the source material at the requisite doctrinal depth, treating legal principles as abstract concepts rather than grounded, citable authorities.
  
  \item \textbf{Proclivity for Irrelevance and Inefficient Content Generation:} A second pervasive issue was the generation of content that substantively drifted from the core legal or factual premises of the question. Graders, colloquially yet pointedly, categorized this as "slop" - digressive text that, while potentially grammatically correct and thematically adjacent, fails to advance a direct answer. This includes lengthy paraphrases of basic legal principles already assumed by the question, speculative explorations of tangential legal scenarios, or the inclusion of boilerplate disclaimers that add no analytical value. This inefficiency not only obscures the relevant answer but also reflects a model's difficulty in performing the crucial task of issue-spotting and prioritization, a skill wherein human examinees are trained to allocate their limited time and space exclusively to the most salient points.
  
  \item \textbf{Inapt Voice, Structure, and Rhetorical Framing:} The third failure mode pertains to the formal and stylistic conventions of professional legal communication. Model responses were frequently characterized by a distinctly "AI-sounding" cadence, often beginning with overly broad, rephrased introductions that lack the incisive tone expected in high-stakes legal writing. A particularly jarring convention noted by graders was the use of meta-framing, such as prefacing an answer with "As an aspiring Advocate on Record..." or "In my capacity as a legal AI...", despite instructions in prompts not to use such commentary. Such framing breaks the professional illusion and reveals the artificial nature of the author. Furthermore, the structural preferences of the models, which often favor long, generalized paragraphs, clash with the exam specific expectations for concise, point wise answers. The models struggled to adopt the succinct, authoritative, and forum-specific voice that human graders associate with a well-trained legal professional, instead defaulting to a more verbose and generically informative prose style.
\end{itemize}

In practice, these are filing-critical defects that attract direct mark deductions. They explain the model–human gap in subjective papers.

% ===================== 7. WHAT EXAMS REVEAL =====================

\section{Conclusion}
This study demonstrates a clear dichotomy in AI capabilities for legal tasks. On objective, multiple-choice examinations, frontier models meet or even surpass historical human pass marks, demonstrating a robust capacity for short-context legal recall and rule application. However, this proficiency does not translate seamlessly to the domain of subjective, long-form writing, where no model could match the performance of a human topper. The critical deficits lie both in knowledge and in execution: a lack of procedural fidelity, imprecise authority handling, and a failure to adopt forum-specific structure.

These findings compel a two-part definition of what it means to be truly "court-ready." First is the \emph{capacity} to recall and apply legal doctrine at scale, a benchmark the strongest systems now meet. The second, and more elusive, aspect is \emph{reliability} under the practical constraints of legal practice, specifically adhering to procedural defaults, maintaining strict authority discipline, and producing work products that align with judicial expectations. Our results indicate that current systems fall short on this second, crucial dimension for apex court practice in India.

In practical terms, this means AI is best deployed as a supportive tool rather than an autonomous practitioner. Systems can efficiently assist with tasks like searching and verifying authorities, checking consistency across drafts, or cross-referencing case details. Tasks that require full drafting, independent citation, or any action carrying legal responsibility remain beyond current capabilities and should be handled with human oversight.

\section*{Limitations}
\label{sec:limitations}
While our work provides stakeholders with greater insights into today's frontier models it has the following limitations:

\begin{itemize}
    \item \textbf{Exam representativeness.} Publicly available exams serve as proxies for live courtroom practice and may not fully capture real-world complexity.
    \item \textbf{Year coverage \& answer keys.} Our results depend on the official answer keys and the subset of years with high-quality scans, which may limit generalizability.
    \item \textbf{Model selection.} We evaluate a triad of models for AoR 2023 constrained by bandwidth and resources; other models may yield different outcomes.
    \item \textbf{Endpoint variance.} Closed-model updates and run-to-run variability can shift results; we snapshot models at evaluation time.
    \item \textbf{Blind grading sample size.} Five assessors of record provide robust signals, but cannot exhaust all stylistic or interpretive variance.
\end{itemize}
% ===================== REFERENCES =====================
%\bibliographystyle{acl_natbib}
\bibliography{custom}

% ===================== APPENDIX (does not count toward page limit) =====================
\appendix

% ===================== APPENDIX: PROMPT TEMPLATES =====================
\section{Prompt Templates and Interfaces}
\label{app:prompts_all}

\paragraph{Chat template (all models).}
We use a simple two–message chat format across models; only exam-specific prompts differ.
\begin{lstlisting}[style=promptjson]
[
  { "role": "system", "content": "<system_prompt>" },
  { "role": "user",   "content": "<user_message>" }
]
\end{lstlisting}
Objective exams require strict JSON outputs; subjective exams are long-form. (Full exam-specific prompts below.)

%                             -
\section{Objective Exam Prompts}
\label{app:prompts_obj}

\subsection*{CLAT PG (objective)}
\begin{lstlisting}[style=promptjson]
system_prompt:
{
  "You are an aspiring law student taking the Common Law Admission Test (CLAT)
   for Post Graduate programs. You will output ONLY strict JSON objects with your
   answer. Analyze each question carefully and choose the best answer from the
   given options. MARKING SCHEME: Each question carries 1 mark with negative
   marking of 0.25 for wrong answers. You can choose to SKIP a question if you're
   unsure to avoid negative marking. Do not include code fences or commentary."
}

user_message:
{
  "You are taking the Common Law Admission Test (CLAT) for Post Graduate programs.
   Read the question and options and choose one answer. Always return ONLY a JSON
   object with keys: 'answer_label' and 'explanation'. The 'answer_label' MUST be
   one of: 'A','B','C','D' for selecting an option, or 'SKIP' to avoid negative
   marking if unsure. Keep the explanation concise (1-3 sentences). Do not include
   any other keys or commentary.\n
   Question: {question}\n
   Options: {options}\n
   Return JSON now."
}
\end{lstlisting}

\subsection*{CLAT UG (objective)}
Same interface as CLAT PG; the allowed labels are \texttt{A,B,C,D,SKIP} and the exam name is changed to CLAT UG.

\subsection*{DJS / DHJS Preliminary (objective)}
\begin{lstlisting}[style=promptjson]
system_prompt:
{
  "You are a knowledgeable legal expert taking the Delhi Judicial Service Examination.
   You will output ONLY strict JSON objects with your answer. Do not include code
   fences or commentary."
}

user_message:
{
  "You are taking the Delhi Judicial Service (DJS) Examination. This exam tests
   knowledge of Indian law, judicial aptitude, general knowledge and current
   affairs. You are answering questions from: {paper_name}\n
   Instructions:\n
   - Each correct answer carries 1 mark.\n
   - Each incorrect answer carries negative 0.25 mark.\n
   - Each skipped answer carries 0 mark.\n
   - Choose the most appropriate answer based on Indian law and legal principles.\n
   - Return ONLY a JSON object with 'answer_label' (1,2,3,4 or SKIP) and
     'explanation' (brief legal reasoning).\n
   - Do not automatically skip GK/current-affairs questions.\n
   - Skip if unsure to avoid negative marking.\n
   Question: {question}\n
   Options: {options}"
}
\end{lstlisting}

\section{Objective JSON schema (enforced at parse time).}
\begin{lstlisting}[style=promptjson]
{
  "answer_label": "A|B|C|D|SKIP"   
  // or "1|2|3|4|SKIP"             
  "explanation": "1-3 sentences"
}
\end{lstlisting}

%                             -
\section{AoR (Subjective) Prompts}
\label{app:prompts_aor}

\subsection*{Practice \& Procedure (AoR Paper I)}
\begin{lstlisting}[style=prompttext]
"You are taking the Advocate on Record (AOR) Examination.
 You are answering questions for the section Practice and Procedure of the Supreme
 Court of India.

 Important Instructions:
 - Be verbose but keep the marks for the question in mind.
 - Write like a candidate would; do NOT reveal that you are an LLM.
 - Do not include code fences or meta commentary.
 - Provide comprehensive answers; include relevant case law and statutory provisions.
 - Structure your answer logically with clear headings.
 - Be precise and accurate in legal terminology."
\end{lstlisting}

\subsection*{Drafting (AoR Paper II)}
\begin{lstlisting}[style=prompttext]
"You are taking the Advocate on Record (AOR) Examination.
 You are answering questions for the section Drafting.

 Important Instructions:
 - Write like a candidate would; do NOT reveal that you are an LLM.
 - The question carries 20 marks.
 - You will be given context and appendices.
 - Draft the required legal document as specified in the question.
 - Follow proper legal drafting format and structure.
 - Include all necessary components mentioned in 'INSTRUCTIONS'.
 - Use appropriate legal language and terminology."
\end{lstlisting}

\subsection*{Advocacy \& Professional Ethics (AoR Paper III)}
\begin{lstlisting}[style=prompttext]
"You are taking the Advocate on Record (AOR) Examination.
 You are answering questions for the section Advocacy and Ethics.

 Important Instructions:
 - Be verbose but respect the marks allotted.
 - Write like a candidate would; do NOT reveal that you are an LLM.
 - Do not include code fences or meta commentary."
\end{lstlisting}

\subsection*{Leading Cases (AoR Paper IV)}
\begin{lstlisting}[style=prompttext]
"You are taking the Advocate on Record (AOR) Examination.
 You are answering questions for the section Leading Cases of India.

 Important Instructions:
 - Be verbose but respect the marks allotted.
 - Write like a candidate would; do NOT reveal that you are an LLM.
 - Do not include code fences or meta commentary."
\end{lstlisting}

\section{AoR Drafting (Paper II): Exclusion Rationale and Pilot Grader Note}
\label{app:aor_drafting_note}

\noindent\textbf{Rationale.} AoR Drafting is \emph{format-critical}: cause title, party array, prayer, affidavits, signatures/verification, pagination/lineation, margining, and citation form must match Supreme Court templates (per \emph{Supreme Court Rules, 2013}, Order~IV and allied provisions) \citep{scrules2013}. Autoregressive, text-only LLMs without document-template tooling frequently violate these formal requirements even when the narrative is legally plausible. To avoid scoring noise dominated by page-layout compliance, we exclude Drafting from quantitative comparisons and focus on the three reasoning-centric papers (Practice \& Procedure; Advocacy/Professional Ethics; Leading Cases).

\medskip
\noindent\textbf{Pilot grading (one draft).} A certified AoR graded a single LLM Drafting response and marked it “\emph{not evaluable}” due to pervasive formal defects. Representative issues (verbatim categories from the grader):
\begin{itemize}
  \item Missing or malformed \emph{cause title} and party array; prayer block not in prescribed order.
  \item Incorrect or absent references to relevant \emph{Supreme Court Rules}; wrong order numbers.
  \item Affidavit/verification, Vakalatnama, and signature blocks omitted or mispositioned.
  \item Pagination/line numbers and margining absent; citations inconsistently formatted.
  \item One court-fee statement incorrect for SLP (Crl).
\end{itemize}
The full note is archived with the anonymized script (available to reviewers on request). The other three papers were graded under the blinded protocol described in the main text.

\section{AoR Grader Packet \& Instructions}
\label{app:aor_grader_instructions}

\paragraph{Materials provided.}
(1) AoR 2023 question paper (with official marks per question); 
(2) \emph{Answer Script A} (AI-generated, anonymized); 
(3) \emph{Answer Script B} (human-written topper, anonymized).
Graders are not told which script is human or AI.

\paragraph{How to evaluate (high-level).}
Use the official question paper to guide marking and apply the same standards used in real AoR evaluation.
Award marks \emph{per question} out of the official maximum (e.g., a 20-mark question must receive $0$–$20$).
Provide short notes where relevant and an overall comment per script.

\paragraph{Rubric dimensions.}
\begin{itemize}
  \item \textbf{Accuracy of law \& reasoning:} Are principles stated correctly and applied to facts?
  \item \textbf{Case law \& statutes:} Verify that cited authorities exist and are relevant; deduct for fabricated or incorrect citations.
  \item \textbf{Structure \& coherence:} Clear issue $\rightarrow$ rule $\rightarrow$ application $\rightarrow$ conclusion flow; forum-appropriate organization.
  \item \textbf{Depth of analysis:} Beyond surface points; counter-arguments/nuances addressed where pertinent.
  \item \textbf{Language \& expression:} Clear, professional, and appropriate for a Supreme Court exam answer.
\end{itemize}

\paragraph{Partial credit.}
Award partial credit wherever reasoning is substantively sound even if incomplete or imperfectly expressed.

\paragraph{Output expected from graders.}
\begin{itemize}
  \item Question-wise marks (out of the official marks allotted).
  \item Brief evaluator notes (e.g., “case not found,” “well-structured,” “analysis shallow”).
  \item A 2–3 sentence overall comment on the paper’s quality.
\end{itemize}

\paragraph{Note on Drafting (Paper II).}
Drafting is not part of the quantitative comparison in this study due to strict layout/form requirements.
One pilot draft was graded and deemed “not evaluable” owing to pervasive formal defects; see Appendix~\S\ref{app:aor_drafting_note} for the summary note.

\begin{table*}[t]
\centering
\tiny
\setlength{\tabcolsep}{1pt}\renewcommand{\arraystretch}{0.85}
\begin{tabular*}{\textwidth}{l @{\extracolsep{\fill}} *{15}{c}}
\toprule
\textbf{Year} &
\rotatebox{90}{Gemini 2.5 Pro} &
\rotatebox{90}{GPT 5 Chat} &
\rotatebox{90}{DeepSeek R1} &
\rotatebox{90}{DeepSeek Chat v3} &
\rotatebox{90}{Claude Sonnet 4} &
\rotatebox{90}{Mistral Medium 3.1} &
\rotatebox{90}{Qwen 3 235B} &
\rotatebox{90}{Llama 3.3 70B} &
\rotatebox{90}{GPT 4.1 Mini} &
\rotatebox{90}{Gemma 3 27B} &
\rotatebox{90}{Qwen 2.5 7B} &
\rotatebox{90}{Gemma 3 12B} &
\rotatebox{90}{GPT 3.5 Turbo} &
\rotatebox{90}{Llama 3.1 8B} &
\rotatebox{90}{Exam Topper} \\
\midrule
2008 & 145.75 & 134.25 & 136.5 & 136 & 136.75 & \textbf{147} & 136.75 & 132.75 & 125 & 119.25 & 103.5 & 114 & 109 & 99.75 & -- \\
2009 & \textbf{179.25} & 166.75 & 171.25 & 169.25 & 175 & 165.75 & 165.25 & 155.5 & 149.25 & 144.5 & 116 & 117.5 & 114.5 & 97 & 175 \\
2010 & \textbf{167} & 157.75 & 162.75 & 159.25 & 158.25 & 154 & 159 & 151.25 & 142.75 & 139 & 129 & 137 & 128 & 112 & 165 \\
2011 & \textbf{173.5} & 146.75 & 153.25 & 151.75 & 153.75 & 138 & 141 & 125.75 & 129.75 & 115.5 & 89.25 & 108.75 & 97.75 & 96.75 & 173 \\
2012 & 154.75 & 138.75 & 146.25 & 144.75 & 152.75 & 149.75 & 147.75 & 120.5 & 128.5 & 109.75 & 96.5 & 110.5 & 104 & 80 & \textbf{159} \\
2013 & \textbf{168.5} & 151.25 & 164.25 & 166 & 146.25 & 149 & 150 & 135.25 & 133.75 & 132.5 & 111.5 & 114.75 & 107.75 & 97 & 160.75 \\
2014 & \textbf{185.25} & 147.75 & 166.5 & 172.5 & 175 & 169 & 164.25 & 151.5 & 135.25 & 132.5 & 114 & 119 & 115.25 & 103.25 & 171.75 \\
2015 & \textbf{156.25} & 117 & 133 & 131 & 130 & 114.75 & 119.5 & 98 & 91 & 100 & 83.25 & 86 & 74 & 30.75 & 143.75 \\
2016 & \textbf{187.75} & 164 & 178.25 & 165.5 & 174 & 166.5 & 169.75 & 151.75 & 145.75 & 140.5 & 113.75 & 116 & 118 & 91.5 & 174.5 \\
2018 & \textbf{180.25} & 144 & 156.75 & 149 & 150.5 & 147.5 & 143.5 & 130.5 & 131.25 & 138.75 & 116.5 & 112.75 & 106.75 & 102.25 & 159 \\
2019 & \textbf{184} & 145.5 & 163.25 & 162.5 & 165 & 146.75 & 154.75 & 130.5 & 132.25 & 127.75 & 104.75 & 115.75 & 94.75 & 98.5 & 177.25 \\
2021 & \textbf{136.25} & 126.25 & 123.75 & 123.75 & 121.5 & 115 & 115 & 110.75 & 103.75 & 100.75 & 92.75 & 83.25 & 101.25 & 92.25 & 125.5 \\
2022 & \textbf{137.5} & 116.5 & 125.75 & 116.5 & 127.25 & 121.5 & 122 & 114.5 & 106.5 & 107.75 & 88 & 101 & 87.75 & 95.5 & 121 \\
2023 & \textbf{130.5} & 111.75 & 84 & 114.25 & 123 & 120.5 & 112 & 95 & 114.25 & 96.5 & 95.25 & 98.5 & 92 & 94.5 & 116.75 \\
2024 & 99.75 & 93.5 & 98.5 & 97.25 & 100 & 96 & 100 & 87.5 & 92.25 & 84 & 84.75 & 83.75 & 87.25 & 77.75 & \textbf{108} \\
2025 & \textbf{107.75} & 92.5 & 96.25 & 99 & 100.25 & 96.25 & 94 & 93.75 & 96.25 & 93 & 82.75 & 85.5 & 80 & 88.25 & 103.5 \\
\bottomrule
\end{tabular*}
\caption{CLAT UG Results.}
\label{tab:clat-ug}
\end{table*}

\begin{table*}[t]
\centering
\tiny
\setlength{\tabcolsep}{1pt}\renewcommand{\arraystretch}{0.85}
\begin{tabular*}{\textwidth}{l @{\extracolsep{\fill}} *{15}{c}}
\toprule
\textbf{Year} &
\rotatebox{90}{Gemini 2.5 Pro} &
\rotatebox{90}{GPT 5 Chat} &
\rotatebox{90}{DeepSeek R1} &
\rotatebox{90}{DeepSeek Chat v3} &
\rotatebox{90}{Claude Sonnet 4} &
\rotatebox{90}{Mistral Medium 3.1} &
\rotatebox{90}{Qwen 3 235B} &
\rotatebox{90}{Llama 3.3 70B} &
\rotatebox{90}{GPT 4.1 Mini} &
\rotatebox{90}{Gemma 3 27B} &
\rotatebox{90}{Qwen 2.5 7B} &
\rotatebox{90}{Gemma 3 12B} &
\rotatebox{90}{GPT 3.5 Turbo} &
\rotatebox{90}{Llama 3.1 8B} &
\rotatebox{90}{Exam Topper} \\
\midrule
2019 & \textbf{85.27} & 81.75 & 76.75 & 75.25 & 75.75 & 76.5 & 76.75 & 72.75 & 71.5 & 66.5 & 48.75 & 60.75 & 56.25 & 34.5 & -- \\
2020 & \textbf{95.5} & 99 & 79.5 & 83 & 76.25 & 79.5 & 58.75 & 76.25 & 65.25 & 52.5 & 40.5 & 44.25 & 51.25 & 47.75 & 72 \\
2021 & \textbf{105} & 98.75 & 96.5 & 90.25 & 92 & 91.75 & 87.25 & 83.25 & 89.25 & 74.5 & 60.25 & 61.5 & 61.75 & 54 & 85.75 \\
2022 & \textbf{106.5} & 102 & 94 & 93.25 & 87.25 & 92.75 & 90.5 & 84 & 77.75 & 73 & 61.75 & 61.75 & 66 & 51.75 & 94 \\
2023 & 103.75 & \textbf{105} & 88 & 87.5 & 92.75 & 91.25 & 87.75 & 81.75 & 70.25 & 70.25 & 67 & 65.5 & 67.75 & 64 & 95.5 \\
2024 & \textbf{111.25} & 110 & 100 & 101.75 & 94.5 & 98.75 & 90 & 90.5 & 81.75 & 76.25 & 66 & 73.75 & 71.5 & 66.25 & 104.25 \\
2025 & \textbf{95} & 90.25 & 93 & 90.5 & 88.25 & 85.5 & 80 & 78.25 & 85.5 & 72 & 62.5 & 62.75 & 68 & 65.25 & 80 \\
\bottomrule
\end{tabular*}
\caption{CLAT PG Results.}
\label{tab:clat-pg}
\end{table*}

\begin{table*}[t]
\centering
\tiny
\setlength{\tabcolsep}{1pt}\renewcommand{\arraystretch}{0.85}
\begin{tabular*}{\textwidth}{l @{\extracolsep{\fill}} *{15}{c}}
\toprule
\textbf{Year} &
\rotatebox{90}{Gemini 2.5 Pro} &
\rotatebox{90}{GPT 5 Chat} &
\rotatebox{90}{DeepSeek R1} &
\rotatebox{90}{DeepSeek Chat v3} &
\rotatebox{90}{Claude Sonnet 4} &
\rotatebox{90}{Mistral Medium 3.1} &
\rotatebox{90}{Qwen 3 235B} &
\rotatebox{90}{Llama 3.3 70B} &
\rotatebox{90}{GPT 4.1 Mini} &
\rotatebox{90}{Gemma 3 27B} &
\rotatebox{90}{Qwen 2.5 7B} &
\rotatebox{90}{Gemma 3 12B} &
\rotatebox{90}{GPT 3.5 Turbo} &
\rotatebox{90}{Llama 3.1 8B} &
\rotatebox{90}{Exam Topper} \\
\midrule
2011 & \textbf{151.75} & 133.75 & 123.75 & 111.25 & 121.25 & 125 & 92.5 & 95 & 122.5 & 100 & 82.5 & 82.5 & 56.25 & 68.75 & -- \\
2014 & \textbf{178.75} & 156.25 & 140 & 127.5 & 145 & 145 & 131.5 & 115 & 125 & 112.5 & 85 & 87.5 & 78.75 & 73.5 & -- \\
2015 & \textbf{191.25} & 178.75 & 157.5 & 145 & 168.75 & 165 & 138.75 & 131.25 & 145 & 128.75 & 78.75 & 106.25 & 103.75 & 86.75 & -- \\
2017 & \textbf{191.25} & 178.75 & 161.25 & 147.5 & 160 & 153.75 & 123.75 & 122.5 & 131.25 & 120 & 102.5 & 111.25 & 91.25 & 76.75 & -- \\
2018 & \textbf{178.75} & 162.5 & 143 & 135 & 143 & 132.5 & 125 & 123.75 & 111.25 & 108.75 & 66.75 & 91.75 & 87 & 89.5 & -- \\
2019 & \textbf{170} & 155 & 129.75 & 132.25 & 136 & 145 & 116.25 & 112.5 & 98.75 & 100 & 56 & 83.5 & 66 & 78.55 & -- \\
2022 & \textbf{185} & 145 & 148.75 & 136.25 & 151.25 & 147.5 & 131.5 & 108.75 & 120 & 97.5 & 72.5 & 83.75 & 88.75 & 87.75 & -- \\
\bottomrule
\end{tabular*}
\caption{Delhi Judicial Services (DJS) Results. Topper Marks are not released.}
\label{tab:djs}
\end{table*}

\begin{table*}[t]
\centering
\tiny
\setlength{\tabcolsep}{1pt}\renewcommand{\arraystretch}{0.85}
\begin{tabular*}{\textwidth}{l @{\extracolsep{\fill}} *{15}{c}}
\toprule
\textbf{Year} &
\rotatebox{90}{Gemini 2.5 Pro} &
\rotatebox{90}{GPT 5 Chat} &
\rotatebox{90}{DeepSeek R1} &
\rotatebox{90}{DeepSeek Chat v3} &
\rotatebox{90}{Claude Sonnet 4} &
\rotatebox{90}{Mistral Medium 3.1} &
\rotatebox{90}{Qwen 3 235B} &
\rotatebox{90}{Llama 3.3 70B} &
\rotatebox{90}{GPT 4.1 Mini} &
\rotatebox{90}{Gemma 3 27B} &
\rotatebox{90}{Qwen 2.5 7B} &
\rotatebox{90}{Gemma 3 12B} &
\rotatebox{90}{GPT 3.5 Turbo} &
\rotatebox{90}{Llama 3.1 8B} &
\rotatebox{90}{Exam Topper} \\
\midrule
2013 & \textbf{151.75} & 133.75 & 123.75 & 111.25 & 121.25 & 125 & 92.5 & 95 & 122.5 & 100 & 82.5 & 82.5 & 56.25 & 68.75 & -- \\
2017 & \textbf{130} & 112.5 & 102.5 & 72.5 & 105 & 97.5 & 80 & 91.25 & 91.25 & 86.25 & 70 & 80 & 75 & 54.5 & -- \\
2019 & \textbf{145} & 128.75 & 120.5 & 112.5 & 118.75 & 117.5 & 90 & 90 & 96.25 & 85 & 60 & 83.75 & 63.75 & 58.5 & -- \\
2022 & \textbf{143.75} & 127.5 & 113.75 & 111.25 & 115 & 106.5 & 77.5 & 91.25 & 86.25 & 73.75 & 47.5 & 62.5 & 56.25 & 50.5 & -- \\
2023 & \textbf{138.75} & 108.75 & 116.25 & 111.25 & 112.5 & 102.5 & 85 & 91.25 & 86.25 & 78.75 & 50 & 73.75 & 55 & 54.5 & -- \\
\bottomrule
\end{tabular*}
\caption{Delhi Higher Judicial Services (DHJS) Results. Topper Marks are not released.}
\label{tab:dhjs}
\end{table*}

%       Appendix: Official exam contents / syllabus      
\begin{table*}[t]
\centering
\small
\setlength{\tabcolsep}{6pt}\renewcommand{\arraystretch}{1.1}
\begin{tabularx}{\textwidth}{l l Y}
\toprule
\textbf{Exam} & \textbf{Modality} & \textbf{Contents (official)} \\
\midrule
CLAT UG & Objective & English Language; Current Affairs / GK; Legal Reasoning; Logical Reasoning; Quantitative Techniques. \\
CLAT PG & Objective & Core LL.B.\ subjects: Constitutional Law; Jurisprudence; Administrative Law; Contract; Torts; Family; Criminal; Property; Company; Public International Law; Tax; Environmental; Labour / Industrial Law. \\
DJS (Prelim.) & Objective & Constitution; CPC; CrPC / BNSS; IPC / BNS; Evidence; Contract; Partnership; Arbitration; Specific Relief; Limitation; POCSO; Commercial Courts Act; English / GK. \\
DHJS (Prelim.) & Objective & DJS core plus commercial/statutory expansion: TPA; Sale of Goods; Negotiable Instruments; Succession / Hindu laws; Prevention of Corruption; POCSO; SARFAESI / DRT; Labour laws; Commercial Courts; IT; IPRs; English / GK. \\
AoR (SC) & Subjective & Four descriptive papers: Practice \& Procedure of the Supreme Court; Drafting; Advocacy \& Professional Ethics; Leading Cases (official case list). \\
\bottomrule
\end{tabularx}
\caption{Appendix syllabus/contents (official sources: Consortium of NLUs for CLAT; Delhi High Court for DJS/DHJS; Supreme Court of India for AoR) \citep{clat_ug_syllabus_2026,clat_pg_pattern,djs_instr_2023,dhjse_imp_instr,aor_sci_official}.}
\label{tab:appendix-syllabus}
\end{table*}

%       Appendix: Exact years used per exam (this study)      
\begin{table*}[t]
\centering
\small
\setlength{\tabcolsep}{6pt}\renewcommand{\arraystretch}{1.1}
\begin{tabularx}{\textwidth}{l c Y c c}
\toprule
\textbf{Exam} & \textbf{Mod.} & \textbf{Years included (official paper + official key)} & \textbf{Qs/exam} & \textbf{Total MCQs} \\
\midrule
CLAT UG & Obj. & 2008, 2009, 2010, 2011, 2012, 2013, 2014, 2015, 2016, 2018, 2019, 2021, 2022, 2023, 2024, 2025 \ (2017 and 2020 excluded due to key issues). & 200/150/120 & \textbf{3{,}154} \\
CLAT PG & Obj. & 2019, 2020, 2021, 2022, 2023, 2024, 2025 & 120 & \textbf{814} \\
DJS (Prelim.) & Obj. & 2011, 2014, 2015, 2017, 2018, 2019, 2022 & 200 & \textbf{1{,}400} \\
DHJS (Prelim.) & Obj. & 2013, 2017, 2019, 2022, 2023 & 150 & \textbf{850} \\
AoR (SC) & Subj. & 2023 (blinded grader study) &  - &  - \\
\bottomrule
\end{tabularx}
\caption{Appendix year coverage used in this study (objective papers require both official paper and official answer key). AoR is fully descriptive (no MCQs).}
\label{tab:appendix-years}
\end{table*}

\section{Comprehensive Result Matrix Across LLMs and Years}
\label{app:results}
We present the results of LLM Evaluations for CLAT UG, CLAT PG, DJS and DHJS in Table \ref{tab:clat-ug}, Table \ref{tab:clat-pg}, Table \ref{tab:djs} and Table \ref{tab:dhjs} respectively.

\section{Ethical Considerations}

This research involved human evaluation of high-stakes professional materials, guided by the following ethical protocols:

\begin{itemize}
    \item \textbf{Voluntary Expert Participation:} Certified Advocate-on-Record (AoR) evaluators participated on a voluntary basis. Their involvement was motivated by a professional interest in advancing understanding of technology within the legal field, and their contribution is gratefully acknowledged.
    
    \item \textbf{Managed Workload and Anonymization:} To respect the time of our volunteer experts, the evaluation workload was carefully limited to a manageable number of anonymized scripts. This prevented fatigue and ensured the integrity of the subjective assessment process.
    
    \item \textbf{Blinded Evaluation for Objectivity:} A paired, blind methodology was employed, making certain that evaluators were unable to differentiate between scripts created by humans and those by AI. This was essential in reducing bias and achieving unbiased and equitable comparisons. However, the stylistic disparities between human work and that of LLMs often resulted in comments suggesting suspicion that the responses might have been AI-generated.
    
    \item \textbf{Integrity and Transparency:} The study uses only officially released public materials to avoid compromising exam integrity. We transparently report both model capabilities and their significant limitations in procedural and drafting fidelity, emphasizing the continued necessity of human oversight in legal practice.
\end{itemize}

\end{document}